# On chip scalable highly pure and indistinguishable single photon sources in ordered arrays: Path to Quantum Optical Circuits

*Jiefei Zhang,*[\*,†, ‡,⊥] *Swarnabha Chattaraj,*[†, §] *Qi Huang,*[†, ⊥] *Lucas Jordao,*[‡] *Siyuan Lu,*[§] *and Anupam Madhukar*[\*,‡,⊥,§]

[\*]Corresponding Authors

[†]These authors contributed equally

[‡]Department of Physics and Astronomy, University of Southern California, Los Angeles, California 90089, USA

[⊥]Mork Family Department of Chemical Engineering and Materials Science, University of Southern California, Los Angeles, California 90089, USA

[§]Ming Hsieh Department of Electrical Engineering, University of Southern California, Los Angeles, California 90089, USA

[§]IBM Thomas J. Watson Research Center, Yorktown Heights, New York, 10598, USA

**ABSTRACT:** Realization of quantum optical circuits is at the heart of quantum photonic information processing. A long-standing obstacle however has been the absence of a platform of single photon sources (SPSs) that *simultaneously* satisfies the following required characteristics: spatially ordered SPS arrays that produce, on-demand, highly pure, and indistinguishable single photons with sufficiently uniform emission characteristics across the array, needed for controlled interference between photons from distinct sources to enable functional quantum optical networks. Here we report on such a platform of SPSs based upon a novel class of epitaxial quantum dots. Under resonant excitation, the SPSs (without Purcell enhancement) show single photon purity >99% ($g^{(2)}(0) \sim 0.015$), high two-photon Hong-Ou-Mandel interference visibilities of 0.82±0.03 (at 11.5K), and spectral nonuniformity <3nm - within established locally tunable technology. Our platform of SPSs paves the path to creating on-chip scalable quantum photonic systems.

On-chip quantum optical circuits aimed at addressing communication[1], cryptography[2], simulations[3,4], sensing[5], and computation[6] invariably require spatially-ordered arrays of single or heralded-pair of photon emitters surrounded by a photon emission rate control and emitted photon directional guiding unit (resonant cavity / nanoantenna, waveguide)[7,8], a functional element to control the relative phase of photons between paths, beam combiners (directional couplers) single photon detectors in co-designed arrays (Fig.1). Currently the physical realization of such quantum photonic systems is limited by the lack of suitable spatially ordered arrays of deterministic on-demand single photon emitters that also simultaneously satisfy the requirements of brightness, near unity single photon purity[9] and indistinguishability[10], and emit within locally tunable range of wavelengths across a scalable array. In this paper, we present a class of single photon emitters that show considerable promise of satisfying all the above noted requirements and thus pave the way to the first step towards on-chip network of controllably connected single photons to enable interference and entanglement[11] between two or more photons from known distinct sources.

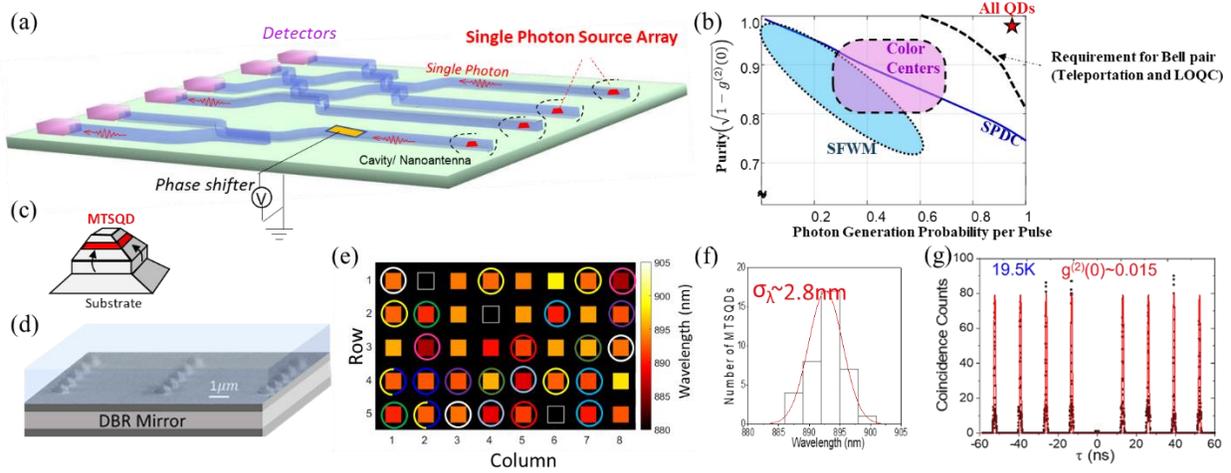

**Figure 1**. (a) Schematic of an on-chip quantum optical circuit-illustrating the need for single photon sources (marked as red dots) to be in spatially ordered locations to be interconnected in a network via light manipulating units and eventually detectors (b) plot of purity vs photon generation probability per pulse for single photon sources. The black dashed line (from Ref. 9) indicates the requirement for 2-photon Bell pair generation using linear optics satisfying the Clauser-Horne-Shimony-Holt-inequality, the minimum requirement for quantum teleportation and for LOQC. The information on the SPDC (spontaneous parametric down conversion), SFWM (spontaneous four wave mixing), color center and QD performance range is taken from Refs 8, 9, 12-15. (c) Schematic of the as-patterned pedestal-shaped mesas on which the QD (red region) is formed selectively during growth using SESRE (d) Schematic composite of the SEM image of the as-formed array of pyramids bearing $In_{0.5}Ga_{0.5}As$ single quantum dot of panel (b) buried by the overgrowth (translucent layer) of the morphology planarizing layer (here GaAs) (e) Color-coded image of the neutral exciton emission wavelength of MTSQDs in the 5×8 array buried under the planarized overlayer. Data collected with 640nm excitation. The black blocks with white outline indicate non-emitting MTSQD pixels. Like-color circles mark MTSQDs emitting within 250μeV. (f) Histogram showing the emission wavelengths of the planarized QD array centered at 893nm with a narrow 2.8nm standard deviation. (g) Histogram of coincidence counts of neutral exciton emission from MTSQDs using HBT setup (Typical photoluminescence emission is shown in Fig. S5 of Supplementary Information) with resonant excitation and with π/2 pulse at 19.5K. The red curve is calculated based on theory[16].

To date the on-chip solid-state quantum emitters that have been extensively employed in investigations of quantum optical phenomena usually attempt to mimic a two-level atomic-like object and may be classified into the categories of semiconductor epitaxial quantum dots (QDs)[7,8], native defect (ND) complex related deep-levels, and implanted defects (ID)[14,15] created with pre-determined choice of implant to provide operational wavelengths of interest for a particular class of functional technology. Of these, the QDs are demonstrably on-demand whereas the NDs and IDs are invariably probabilistic. The quantum emitters reported in this paper are a novel class of quantum dots[17-20] that, as we demonstrate here, satisfy all the above noted requirements and thus are suited to enabling quantum optical circuits[21]. For proper perspective, we note that the epitaxial QDs are formed in a few different ways but unlike the class employed here dubbed substrate-encoded size-reducing epitaxy (SESRE)[22,23], suffer from some combination of lack of adequate spatial ordering, sufficient spectral uniformity, compatibility with horizontal architecture demanded by on-chip systems, and scalability (see Sec.1, SI). The SESRE approach produces size and shape controlled single quantum dot on the top of ordered arrays of in-situ size-reduced pre-patterned nanomesa – dubbed mesa-top single quantum dots (MTSQDs), as depicted (red region) in Fig. 1(c). It is to be noted that the SESRE approach exploits designed *surface-curvature induced stress gradients* to direct preferentially adatom migration from designed crystallographic sidewall planes to the mesa top plane during deposition to achieve site-selective epitaxy. It thus is applicable to lattice matched (e.g. GaAs/AlGaAs) and mismatched (e.g. GaAs/InGaAs) material combinations[22,23]. The pyramidal morphology of the MTSQDs is planarized with further growth of an overlayer as depicted in Fig.1(d). In this work, we report on InGaAs MTSQD 5×8 arrays as discussed next.

The 5×8 array comprises 4.25ML $In_{0.5}Ga_{0.5}As$ MTSQDs grown on GaAs(001) starting square nanomesas with edges oriented along <100> directions and of size~300nm with a pedestal shape (Fig. 1(c))[20]. The pedestal shape enables planarization following MTSQD formation on size-reduced mesa top of ~15nm lateral size[20]. The starting nanomesa arrays reside on GaAs with a suitably designed AlAs/GaAs DBR underneath (see Method) to enhance the photon collection efficiency at the first objective lens by ~10, bringing it to ~12% for optical measurement purpose (Sec 2, Supplementary Information). The position accuracy of the MTSQDs is controlled by the lithographic uncertainty which is ~5-10nm in the present case. The emission spectral nonuniformity ($\sigma_\lambda$) of the synthesized planarized MTSQDs array is found to be 2.8nm (Fig. 1(e) and (f)) with 29 pairs of QD emitting within 250μeV and a set of 6 QDs within 250μeV marked in like-color circles. With established local tuning technology using the Stark effect[24], the entire array can be brought to resonance as needed for creating optical circuits. The observed spectral non-uniformity can be further reduced by reducing the alloy fluctuation amongst MTSQDs[20] (Fig. S2, Supplementary Information) to reach potentially sub-nm scale spectral nonuniformity. Each individual MTSQD has sharp neutral exciton emission and produces highly pure single photons. One typical measured $g^{(2)}(\tau)$ for neutral exciton emission from the MTSQDs (the spectrum shown in Fig. S5(a), Supplementary Information) under resonant excitation (Method and Fig. S3, Supplementary Information) is shown in Fig. 1(g), revealing a $g^{(2)}(0)$ of 0.015 and attendant single photon purity ~99.2% (extracted based on measured data and theory, Sec 4, Supplementary Information). High purity single photon emission is observed in other MTSQDs (Fig. S6, Supplementary Information). The measured single photon purity is consistent with previously reported behavior of MTSQDs[17,18,19,20]. The planarized MTSQDs are thus highly pure (purity >99%) single photon emitters with controlled position and high spectral uniformity (<3nm). Such characteristics provide strong incentive for scaling the 5×8 arrays to order 50×50 as similar characteristics across such arrays will enable movement towards realizing quantum optical circuits and networks containing hundreds to thousands of SPSs needed for large scale quantum communication[25] and quantum simulations for quantum chemistry[26].

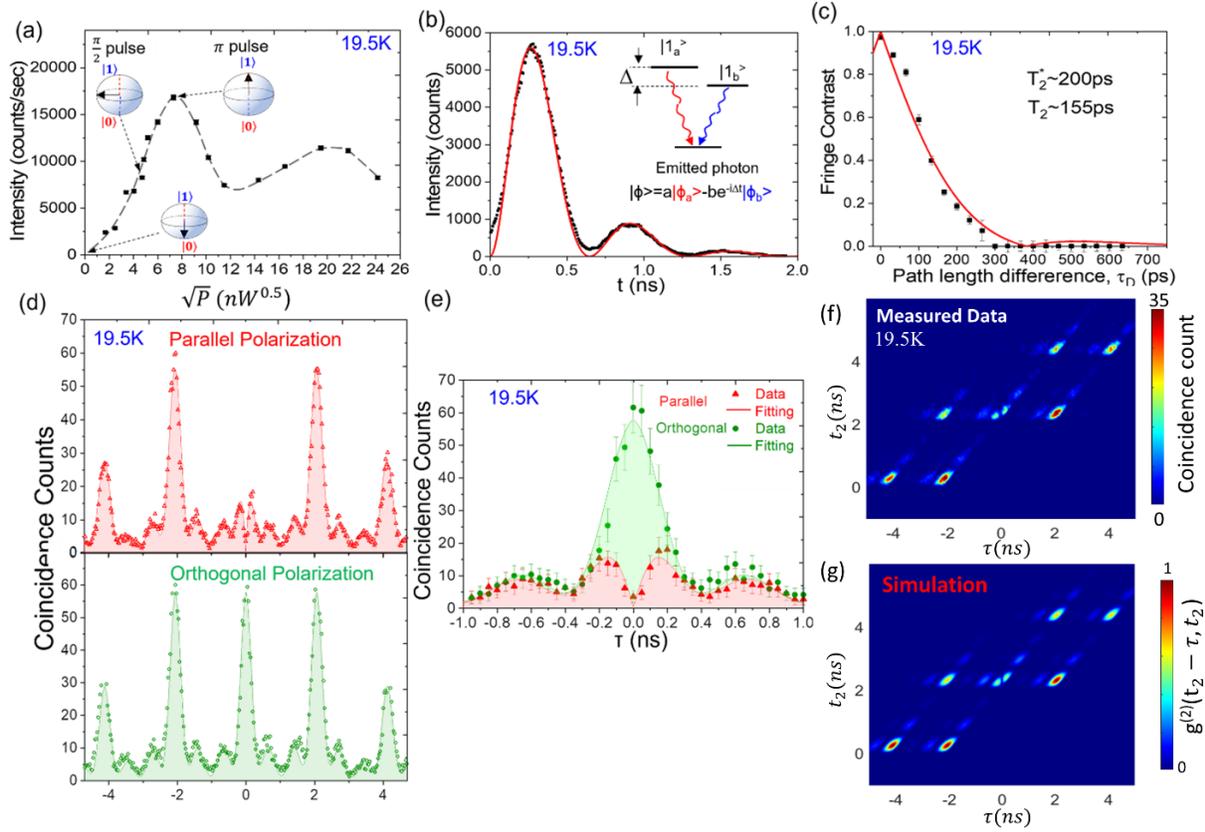

**Figure 2**. **Coherent control of exciton state, photon emission and photon indistinguishability.** (a) Power dependent behavior of peak intensity versus the square root of laser power (proportional to excitation pulse area) from the neutral exciton emission showing clear Rabi oscillation. Error of the measured intensity is within the symbol size. The Bloch sphere inserts represent the switching from ground state $|0\rangle$ (empty QD) to one exciton state $|1\rangle$. (b) Measured time-resolved fluorescence from the MTSQD excited under resonant excitation with π/2 pulse (19.6nW (1.6W/cm$^2$)) at 19.5K. The red curve is fitting to the data using Eq. 1 from our three-level model. The inset shows the three-level diagram of our MTSQDs with Δ the energy separation of the two exciton levels $|1_a\rangle$ and $|1_b\rangle$. The photon emitted from such coherently populated two states is of the form $|\varphi\rangle = a|\varphi_a\rangle - be^{-i\Delta t}|\varphi_a\rangle$ (c) Measured first order interference fringe contrast of photons from the MTSQD as a function of time delay between the two branches of the Michelson interferometer (MI) under resonant excitation with π/2 pulse at 19.5K. (d) Histogram of coincidence counts [τ from -5ns to 5ns] of TPI with data collected under parallel (red triangles, top panel) and orthogonal (green squares, bottom panel) configuration. (e) Histogram of coincidence counts [τ from -1ns to 1ns, covering whole photon wavepacket] of TPI using HOM interferometry with data collected under parallel (red triangles) and orthogonal (green squares) configuration with laser leak corrected. The red and green curves are the fits to the data using Eq. 3 and 4. (f) Measured time-resolved HOM coincidence counts under parallel configuration (top panel). The coincidence counts are plotted as a function of detector 2 detection time ($t_2$) and the time difference between detection event of the two detectors (τ). Panel (g) shows the simulation result using the three-level model based on the known emission dynamics parameters obtained from Fig. 2(b) and (e).

Next, we examine the internal quantum efficiency of the single photon emission and the photon indistinguishability using resonant excitation scheme, as these are the two key important figures-of-merit of SPSs for quantum information processing. The internal quantum efficiency is assessed through power dependent photoluminescence (PL) measurement. Fig. 2(a) shows the measured intensity (*I*) of neutral exciton emission from MTSQD (PL spectrum, see Fig. S5(a) of supplementary) as a function of excitation power (*P*). Well-pronounced Rabi oscillations are observed, indicating coherent manipulation of the QD state between ground state (no exciton, $|0\rangle$), and excited state (one exciton, $|1\rangle$) (shown in the Bloch spheres in Fig. 2(a)). At π pulse, one single exciton is created per pulse. Guided by the detected photon counts at π pulse (17K/sec), the collection efficiency of emitted photon by the first objective lenses of the measurement setup (12±1%) and the detection efficiency of the setup ($\sim 1.81 \times 10^{-3}$), we estimate that one photon is emitted from the MTSQD per excitation pulse. Thus, the internal quantum efficiency of the MTSQD is ~100% (for details see Sec 2, Supplementary Information) indicating good material quality and feasibility of reaching unity probability of photon generation per pulse. Besides the single photon purity and the internal quantum efficiency, the photon indistinguishability is the third key intrinsic figure-of-merit to be addressed in assessing the potential of MTSQD SPSs. We examine the photon emission decay time and coherence time through resonant excitation, as these are the two key timescales controlling two photon interference visibility. Figure 2 (b) shows the measured time resolved PL (TRPL) spectrum from the MTSQD whose integrated emission under the same excitation condition is shown in Fig. S5(a) of supplementary. The TRPL spectrum show oscillatory behavior. This is a temporal beat signal stemming from self-interference of the photon wavepacket on the single-photon detector[27,28,29]. The beats indicate that the emitted single photon is in a coherent superposition of two energy states with an energy separation less than the 15μeV spectral window, suggesting that the MTSQD has a three-level electronic structure shown in the inset of Fig. 2(b). The emitted photon wavepacket coming from $|1_a\rangle$ and $|1_b\rangle$ can be represented as $|\varphi\rangle = a|\varphi_a\rangle - be^{-i\Delta t}|\varphi_a\rangle$. Thus, the time dependent detection probability of the photon wavepacket can be expressed as (details in Sec.7a, Supplementary Information),

$$I(t) \propto \left| e^{-i\Delta t} e^{-\frac{t}{2T_1^{(a)}}} - e^{-\frac{t}{2T_1^{(b)}}} \right|^2 \quad (1)$$

with $\Delta = \omega_a - \omega_b$ and $T_1^{(a)}$ and $T_1^{(b)}$ the radiative decay times. The red curve in Fig. 2(b) shows the fitted result indicating $T_1^{(a)} = T_1^{(b)} = 0.35 ns$ and $\Delta = 6.4 \mu eV$. The time resolved resonance fluorescence thus indicates that there are two energy states separated by 6.4μeV. The two finely separated states are most likely the well-known fine structure split (FSS) states[7,8,19] arising from the loss of $C_{2v}$ symemtry of the confinement potential due to alloy fluctuation, piezoelectric field, and strain. This is also consistant with our previous findings of FSS<10μeV[19]. The dipole orientations of the two FSS states are most likely at an angle to the crystalgraphic [110] and [-1 1 0] directions[19,30,31]. In such a case, when we excite the QD with the excitation laser polarized along [110], both FSS states are pupulated and detected, leading to the observed temperal beating in TRPL. Such behavior is also seen in other MTSQDs with $\Delta = 3.8\mu eV$ and $T_1^{(a)} = T_1^{(b)} = 0.55ns$ (Fig. S7 Supplementray Information).

To establish the coherence time (T$_2$), we carried out single-photon interference measurement in an unbalanced Michelson interferometer (MI). Figure 2 (c) shows the measured 1$^{st}$ order interference fringe contrast as a function of time delay ($\tau_d$) between the two arms of the interferometer. With the emitted photon wavepacket $|\varphi\rangle$ being a coherent superposition of states $|1_a\rangle$ and $|1_b\rangle$, the fringe contrast can be expressed as (details in Sec.7b of Supplementary Information)

$$Fringe\ contrast = \frac{2(1+\Delta^2 T_1^2)}{\Delta^2 T_1^3}\left[\int_0^\infty dt\ e^{-\frac{t}{T_1}}\sin\left(\frac{\Delta}{2}t\right)\sin\left(\frac{\Delta}{2}(t+\tau_d)\right)\right]e^{-\frac{\tau_d}{2T_1}-\frac{\tau_d}{T_2^*}} \quad (2)$$

From the fitting using Eq.2 (red curve in Fig. 2(c)) we obtain a dephasing time $T_2^* \sim 200ps$ and coherence time T$_2$~155ps. Equation 2 also predicts a beating pattern in the fringe contrast resulting from the self-interference of the photon from the two exciton states[27]. At $T_2^* \sim 200ps$, an expected second peak of fringe contrast should emerge at delay $\tau_d$~450ps with a fringe contrast of 2%, which is beyond our instrument's capability and thus not observed. The study of decay lifetime and photon coherence time indicates that the MTSQDs provide highly pure single photons in coherent superposition of two finely split energy states.

The photon indistinguishability is studied through two photon interference (TPI) using Hong-Ou-Mandel (HOM) interferometer (Fig. S3 of supplementary and Methods). Figure 2 (d) and (e) shows the measured coincidence counts between the two output ports of HOM interferometer as a function of time difference ($\tau$) between consecutive detection events in the co-polarized (parallel, upper panel) and cross-polarized (orthogonal, lower panel) configurtions where the λ/2-waveplate rotates the polarization of photons by 90° in one branch to make the two photons intentionaly distinguishable. The data have been corrected for the leakage of pulsed excitation laser under the measurement condition. The two photon interference at τ=0 at the second beam splitter generating energy-entangled photon pairs is manifested in the strong reduction of the coincidence counts at τ=0 in the case of co-polarized configuration as compared to the data taken in the cross-polarized configuration (seen in the expanded view shown in Fig. 2(e)). The beating pattern accuring at ±0.6ns from the main peaks at τ=0, ±2ns and ±4ns in the measured HOM coincidence data (Fig. 2(d)) further confirms that the emitted photon wavepacket is in a coherent superposition of the two states separeted by 6.4μeV as extracted from the time-resolved PL data (Fig. 2 (b)). The observed beat signal originates from single-photon self-interference, not two-photon interference.

Using the total counts at the central peak (covering a range of -1ns to 1ns, Fig. 2(e)) for parallel ($C_\parallel$) and orthogonal ($C_\perp$) polarization, we calculate the TPI visibility[8,28] $V = (C_\perp - C_\parallel)/C_\perp$ = 0.55±0.025. After correcting for the non-zero probability of double excitation of the QDs ($g^{(2)}(0)$ ~ 0.015 (Fig. 1(f)), we find the TPI visibility to be $V_c$~0.57±0.025. The TPI visibility is comparable to the best reported values on QDs in samples without Purcell effect at 19.5K[8]. The HOM coincidence counts data (Fig. 2 (d) and (e)), besides providing a value for the TPI visibility, also allow extracting $T_2^*$ of the photons. With photons from MTSQD being in a coherent superposition of two finely split energy states, the HOM coincidence counts $g_\parallel^{(2)}(\tau)$ and $g_\perp^{(2)}(\tau)$ can be respresented as (details in Sec 7.b of Supplementary Information),

$$g_\parallel^{(2)}(\tau) = \int_0^\infty dt\, e^{-\frac{2t}{T_1}} \sin^2\left(\frac{\Delta}{2}t\right) \sin^2\left(\frac{\Delta}{2}(t+|\tau|)\right) \left[1 - e^{-\frac{2|\tau|}{T_2^*}}\right] e^{-\frac{|\tau|}{T_1}} \quad (3)$$

$$g_\perp^{(2)}(\tau) = \int_0^\infty dt\, e^{-\frac{2t}{T_1}} \sin^2\left(\frac{\Delta}{2}t\right) \sin^2\left(\frac{\Delta}{2}(t+|\tau|)\right) e^{-\frac{|\tau|}{T_1}} \quad (4)$$

We use Eqs. 3 and 4 to analyze the measured data with the known decay lifetime $T_1$ and Δ obtained from the time-resolved PL data shown in Fig. 2(b). The dephasing time $T_2^*$ is obtained using maximum likelihood meathod[32] with the instrument response function folded into the fitting. We obtain $T_2^*$~0.58ns through fitting (red curves, upper panel of Fig. 2(d) and panel (e)). Such a long dephasing time is also clearly seen in the *time-resolved* HOM coincidence count plot shown in Fig. 2(f) where the coincidence counts ($\propto g^{(2)}(t_1, t_2)$) corresponding to photon detection at time $t_1$ and $t_2$ at the two detectors are plotted as a funciton of $t_2$ and $\tau$ ($= t_2 - t_1$). The calculated HOM $g^{(2)}(t_1, t_2)$ using our three-level model and $T_2^*$ of 0.58ns is shown in the panel of Fig. 2 (g). The calculated result matches the experimental data, supporting the validity of the three-level model and futher supporting the inference of dephasing time $T_2^*$ of 0.58ns for the MTSQDs.

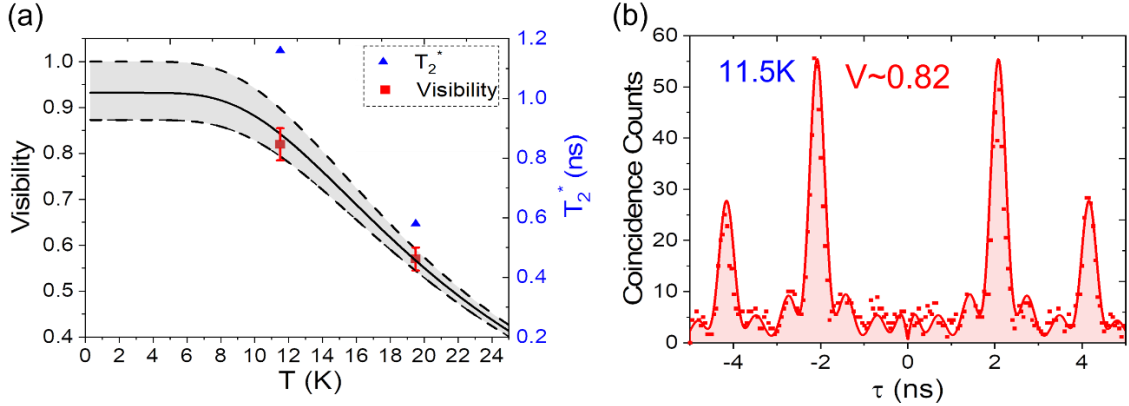

**Figure 3. Highly indistinguishable photon emission and effect of phonon.** (a) Measured temperature dependent TPI visibility (photon indistinguishability) and dephasing time of the measured MTSQD (red dot for TPI visibility, blue triangle for dephasing time). The calculated TPI visibility using phonon dephasing time reported in Ref. 33 for InGaAs/GaAs material system is shown as black line. The grey region marks the range of TPI visibility as a function of temperature given the error range of measured TPI visibility data at 19.5K. (b) Histogram of coincidence counts [τ from -5ns to 5ns] of TPI with data collected under parallel configuration with laser leak corrected at 11.5K with π/2 pulse excitation. The red line is fit to the data using Eq. 3.

The dephasing time $T_2^*$ obtianed from the HOM data is comparable to the best reported perfomance of QDs in the literature at 19.5K[33, 34, 35]. The discrepancy between $T_2^*$ obtained from the HOM and obtianed from MI has been previously reported[33, 34] and is attributed to the difference in integration times: ~sec for MI measurements and ~ns for HOM. Consequently, in HOM, photon dephasing from exciton interaction with the acoustic phonon bath (picosecond range) is probed but not the dephasing from interaction with fluctuating electrostatic environment (microsecond range)[35, 36], leading to a longer $T_2^*$. With resonant excitation cutting down on the spectral diffusion induced dephasing, the observed dephasing time $T_2^*$ and TPI visibility is limited by dephasing process induced by acoustic phonons. The TPI visibility as a function of temeprature can be written as[33] $V(T) = \frac{\Gamma}{\Gamma_{SD}+\gamma(T)+\Gamma}$ where $\Gamma = 1/2T_1$ , $\gamma(T) = \frac{\gamma_0}{\exp(\frac{\alpha}{T})-1}\left[\frac{1}{\exp(\frac{\alpha}{T})-1}+1\right]$ is the temperature dependent phonon induced dephasing rate and $\Gamma_{SD}$ is the spectral diffusion induced dephasing rate.

Using the known temperature dependence of the exciton dephasing rate in the InGaAs/GaAs material system[33] and the measured TPI visibility data at 19.5K, the calculated TPI visibility as a function of temperature is shown as black line in Fig. 3(a). The two gray lines cover the range of TPI visibility due to the uncertainty in $\Gamma_{SD}$ coming from the possible error of the TPI visibility data at 19.5K. The data indicate that the $\Gamma_{SD}$ is longer than 11ns and the TPI visibility, still without Purcell enhancement, is expected to be ~93% at 4K. As a check on the predicted temperature dependence, we revived a LHe cryostat and measured the photon indistingushability at 11.5K (the base temperature of the LHe cryostat). The measured HOM coincidence counts histogram under parallel configuration is shown in Fig. 3(b). The data indicate an as-measured TPI visibilty of 0.80±0.03 and TPI visibility of 0.82±0.03 corrected for multiphoton events. The red line is the fit to the data using Eq. 3 showing a dephasing time $T_2^*$ of 1.16ns. The measured TPI visibility at 11.5K matches the predicted curve as shown in Fig. 3(a). Accounting for the two measured TPI visibility data, one can extrapolate and predict that the TPI visibility can be ~90% at 4K without Purcell factor (Fig. S8 of supplimentary). With a Purcell factor of 5, one can further improve the TPI visibility to 97% (Fig. S8, Supplimentary Information) to be comparable to the best reported value[37,38]. It can thus be concluded that the MTSQDs can generate single photons with >99% purity and with expected >97% photon indistinguishability when integrated in a cavity for Purcell enhancement and operating at 4K.

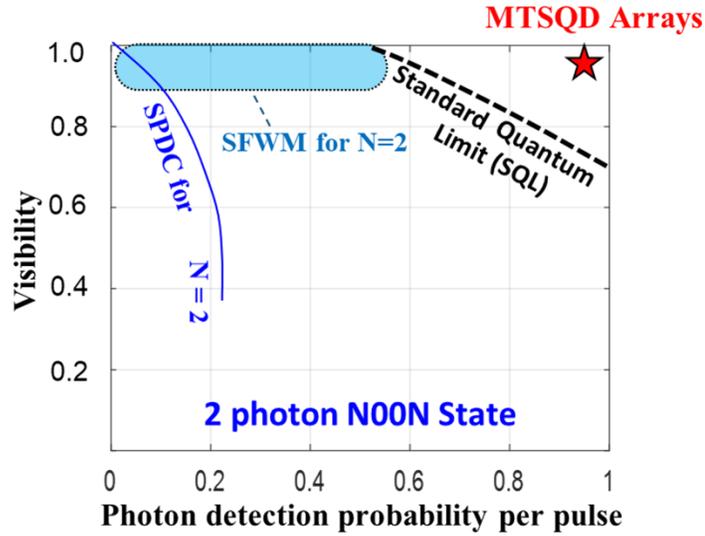

**Figure 4. Potential of MTSQD Arrays for Quantum Sensing.** The requirement for the visibility vs photon detection probability per pulse of single photon sources for generation of 2 photon N00N states for quantum sensing at the standard quantum limit[10]. Unlike the traditional approach of SPDC sources[12] or SFWM sources[13], The MTSQDs in ordered arrays show great potential to satisfy the requirements of near unity visibility and photon detection probability per pulse.

The SESRE based MTSQD arrays provide a (so far, the only) highly promising approach to realizing the needed single photon source platform that satisfy all three of the system level requirements of being in spatially ordered array, having adequate spectral uniformity (nonuniformity within the range of demonstrated on-chip local tuning tethodologies[24,39,40,41]) and also individual function level requirement of simultaneously near unity purity, indistinguishability and internal quantum efficiency. The unique control on position of MTSQDs and their spectral uniformity makes such MTSQD arrays also suitable for realizing multiphoton interference/entanglements. Given that the spectral uniformity of the MTSQDs are within the on-chip tuning range, e.g. ~1-3V tuning voltage using Stark effect to tune wavelength of 3nm[24], one can bring two or more MTSQDs to resonance and to create controlled interference of photons[11] from these different MTSQDs on-chip to create multiphoton entangled states, i.e. N00N states with potentially high efficiency and photon detection probability per pulse for applications in sensing and metrology[10] (Fig. 4) as well as linear optical quantum computing[9]. The SESRE approach is scalable to enable large size MTSQD arrays (containing hundreds of MTSQDs) as photon qubit arrays to enable creating optical circuits and networks for creating a quantum simulator solving quantum problems on chemical reaction dynamics and molecular electronic structure[26]. Such approach also enables a foreseeable path to realize quantum optical circuits containing thousand SPSs for quantum computation[25]. Thus, the demonstrated characteristics of MTSQDs constitute a compelling case for further exploration and rigorous assessment of a potentially viable platform for moving on-chip quantum photonics to the long-awaited next level—creation of well-designed functional quantum optical circuits. To this end, further work is needed on establishing the statistics of photon emission characteristics amongst large arrays of MTSQDs, extending wavelength to 1550nm using InP/InAs material combination as well as establishing control on photon emission rate and directionality through monolithic integration of MTSQDs with light manipulating units using conventional waveguide[7,8,42], 2D Photonic crystal approaches[7,8,40,43] and our newly proposed approach of using Mie resonance of interacting dielectric building block based metastructures co-designed to provide all the needed light manipulation functions[17,21]. Equally important, leveraging Si photonic technology, hybrid integration[44, 45, 46] via transfer printing or flip-chip bonding is also an important direction.

**Methods**

**Sample Fabrication.** The sample studied here contains a buried 5×8 array of MTSQDs sitting on top of a DBR mirror where the DBR is designed to enhance the photon collection efficiency. The DBR mirror is grown with 17 pair of GaAs (65.6nm) and AlAs (78.3nm) at 600°C with $\tau_{GaAs}$=2sec/ML, $\tau_{AlAs}$=2sec/ML and $P_{As}$=3 × $10^{-6}$torr. Nanomesas of size ~300nm with a pedestal shape are created with electron beam lithography and wet chemical etching on the GaAs layer on top of the grown DBR mirror. After growth of 271ML GaAs under condition with $\tau_{GaAs}$=4sec/ML $P_{As}$=1.5 × $10^{-6}$torr at 600°C, the mesa top opening size is reduced from ~300nm to ~20nm during the size-reducing growth on mesa resulting from the surface-curvature induced adatom migration to mesa top. To form a MTSQD on the size-reduced mesa top, 4.25ML $In_{0.5}Ga_{0.5}As$ is deposited at 520°C. A subsequent 1346ML GaAs is grown to cap the QD, and to convert the surface morphology from pinched mesas to near flat surface[20] and have the QD located at the antinodes of the DBR/Air structure, ~280nm away from the DBR/Air interface.

**Measurement Setup.** Optical studies are carried out with sample mounted in the cryostat in vertical excitation and vertical detection geometry as shown in Fig. S1 in supplementary. A Ti-Sa laser is used for optical excitation generating excitation pulses of ~3ps with a repetition rate of 78MHz with laser pulse peak wavelength tuned to the neutral exciton emission wavelength of MTSQDs within accuracy of ±0.01nm. For all measurement (PL, time-resolved PL, HBT, HOM), We excite with laser electric field along [110] direction with accuracy of ±3° and detect photons with polarization along [-1 1 0] direction also with accuracy of ±3°. An extinction ratio >1×$10^7$ is established for resonant excitation studies reported here. The collected photons from the MTSQDs are spectrally resolved with spectrometer of 15μeV resolution and detected by superconducting nanowire detectors that have a time jitter of 50ps.

**HBT.** The photons emitted from the MTSQDs are spectrally filtered by the high resolution spectrometer with 15μeV resolution and enters the 50/50 beamsplitter of the Hanbury-Brown and Twiss setup. Photons passing through the 50/50 beamsplitter are coupled to polarization maintaining fibers through two collimators at the transmitted and reflected ports of the beamsplitter then directed to two superconducting nanowire detectors for detection. The instrument response function for the HBT measurement is ~50-100ps.

**HOM.** For the study of indistinguishability, we use the Ti-Sa laser to generate pairs of excitation pulses of width ~3ps with a time separation Δt=2ns, controlled by an unbalanced Michelson interferometer built on laser side. The impinging laser is also polarized along [110] on the sample. The emitted photons from the MTSQD are polarization discriminated by the polarizer to select emission polarized along [1-10] and is fed through the spectrometer and enters the first 50/50 beam splitter of the Hong-Ou-Mandel interferometer as shown in Fig. 2(d). Photons passing through the beamsplitter are coupled to polarization maintaining fibers through two collimators at the transmitted and reflected port of the beamsplitter. The path length difference is designed to match the time delay of 2ns of the excitation pulse. Photons passed through the fibers interfere at the fiber-inline-beam splitter (50/50) and detected by the superconducting nanowire detectors. A λ/2-waveplate allows for rotating the polarization of photons in one arm of the interferometer with respect to the other arm in order to make the photons distinguishable on purpose for reference measurement. The instrument response function for the HOM measurement is ~50-100ps.

## ASSOCIATED CONTENT

**Supplementary Information**. Detailed information the SESRE approach for MTSQD growth, optical setup for resonant measurement, and theoretical three-level model developed and used for data analysis.

## AUTHOR INFORMATION

**Corresponding Author**

*Jiefei Zhang, orcid.org/0000-0002-7329-3110; Email: jiefeizh@usc.edu

*Anupam Madhukar, orcid.org/0000-0002-1044-1681; Email: madhukar@usc.edu

**Author Contributions**

†These authors contributed equally.

**Competing Interest**

These authors declare no competing interests.

**Notes**

The authors declare no competing financial interest.

ACKNOWLEDGMENT

This work was supported by Air Force Office of Scientific Research (AFOSR), Grant# FA9550-17-01-0353 and US Army Research Office (ARO), Grant# W911NF-19-1-0025.

# Supplementary Information

# On chip scalable highly pure and indistinguishable single photon sources in ordered arrays: Path to Quantum Optical Circuits

*Jiefei Zhang,*[*,†,‡,⊥] *Swarnabha Chattaraj,*[†,§] *Qi Huang,*[†,⊥] *Lucas Jordao,*[⊥] *Siyuan Lu,*[§] *and Anupam Madhukar*[*,‡,⊥]

*Corresponding Author.
†These authors contributed equally.

[‡]Department of Physics and Astronomy, University of Southern California, Los Angeles, California 90089, USA
[⊥]Mork Family Department of Chemical Engineering and Materials Science, University of Southern California, Los Angeles, California 90089, USA
[§]Ming Hsieh Department of Electrical Engineering, University of Southern California, Los Angeles, California 90089, USA
[§]IBM Thomas J. Watson Research Center, Yorktown Heights, New York, 10598, USA

## 1. SESRE: Spatially ordered and spectrally uniform MTSQDs

Of the competing on-chip solid state quantum emitters, the self-assembled quantum dots (SAQDs), being on-demand source, (and owing to the ease of synthesis) have been studied the most[S1] but, given the underlying fundamental physics of lattice mismatch strain-driven spontaneous formation of the defect-free 3D islands during vapor phase deposition[S2, S3] are random in location and nonuniform in size, shape, and composition that manifests in large inhomogeneity (~50nm) in spectral emission in the most commonly employed system of InGaAs on GaAs (001) substrate[S3]. This has severely limited all demonstrations of on-chip SAQD-cavity-waveguide integrated structures to a single SAQD (found through search over a large number) with a few exceptions that employ structures containing two SAQDs[S4]. The need for the number of physical qubits for practical systems for quantum information processing is estimated to range from a few hundred (for simplest quantum chemistry simulations) to millions (for computation) of SPSs[S5]. To overcome this severe limitation, various approaches have been taken to induce site-selective formation of quantum dots[S6, S7, S8] utilizing chemically and structurally pre-patterned substrates. Interested readers can find pertinent information in references S6, S7, S8. While each of these approaches has its merits, none of them meet *simultaneously* all the strict requirements of (i) having one quantum dot per site in a configuration compatible with on-chip integration (i.e. planar / horizontal) architecture, (ii) the photons having sufficiently uniform spectral properties across a *scalable* array, and (iii) applicable to a wide variety of material combinations thereby covering a wide range of spectral emission from the ultraviolet to visible, and near-infrared to mid- and long wavelength infrared regimes. An exception has been the approach of substrate-encoded size-reducing epitaxy (SESRE)[S9, S10] briefly recalled here.

The MTSQDs reported in this manuscript are synthesized using the SESRE approach[S9] that exploits growth on designed non-planar patterned substrates, i.e. patterned structurally such that the tailored surface curvature induces surface stress gradients (capillarity) that direct adatoms during deposition preferentially to mesa tops[S9, S10] for selective incorporation through control on the relative kinetics of adatom incorporation on the contiguous facets present in the designed curvature. For the (001) surface oriented substrates of the tetrahedrally-bonded semiconductors of groups IV, III-V, and II-VI, the <100> edge orientations of square mesas provide four-fold symmetry and thus potentially symmetric migration of adatoms from the sidewalls to the top (Fig.S1(a)). The preferential incorporation at the mesa-top leads to growth-controlled mesa size reduction (Fig. S1(b)), enabling in-situ preparation of contamination-free and defect-free nanomesa of the desired size utilizing homoepitaxy under controlled growth kinetics. A quantum dot (QD) can be formed on the size-reduced nanomesa with crystallographic controlled size-wall and controlled base length and thickness through heteroepitaxy. Thus, SESRE can be used to synthesize QD in spatially ordered arrays with growth controlled QD size and shape, leading to formation of spatially ordered and highly spectrally uniform QDs. What's more, the presence of sufficiently small nanoscale mesa owing to the presence of mesa free sidewalls leads to substantial strain relaxation which enables the accommodation of QD forming material combinations with significant lattice mismatch using SESRE, unlike growth in arrays of pits that restrict the material combination to nearly lattice matched[S7].

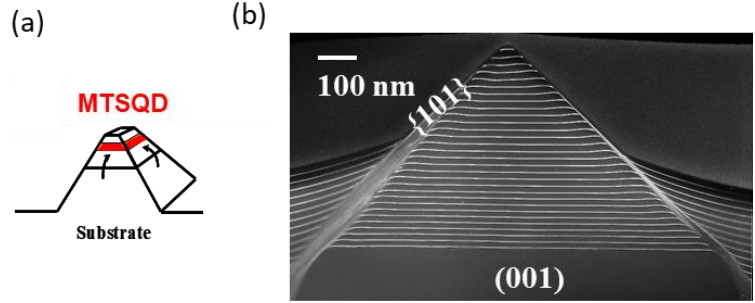

**Figure S1.** (a) Schematic of the MTSQD (red region) forming at apex of each mesa whose spatial selective formation is realized due to the directed atom migration from side of mesa to top (marked by black arrowed) driven by surface-curvature induced stress gradient. (b) Cross-section TEM image of the size-reducing growth on mesa tops with GaAs (dark) and AlGaAs (light, marker layer) [this figure is taken from Ref. S11]

Following SESRE approach, MTSQDs with {103} side walls have been grown and studied[S12, S13]. As-grown non-planarized 5×8 array of 4.25ML $In_{0.5}Ga_{0.5}As$ MTSQDs (Fig. S2(a)) show spectral non-uniformity of ~6-9nm (Fig. S2(b)) centered around 920-930nm from our previous studies[S12, S13]. The planarized 5×8 array of 4.25ML $In_{0.5}Ga_{0.5}As$ MTSQDs (Fig. S2(c)) shows a spectral uniformity of 2.8nm (Fig. S2(d), same as Fig. 1(e)) with a blue shifted center wavelength around 890nm. The thermal annealing during the planarization growth induces intermixing of In and Ga in the MTSQDs that shifts the wavelength and reduces the nonuniformity. The observed spectral uniformity is largely limited by the alloy fluctuation. By reducing the alloy fluctuation, we find that the as-grown non-planarized 5×8 array of binary InAs MTSQDs have an unprecedented uniformity of 1.8nm (Fig. S2(e)) with clusters of 6-7 QDs emitting within 250μeV[S13]. Thus, one expects to have spectral uniformity significantly less than 1.8nm for planarized InAs MTSQDs. All the data present here are from as-grown sample without any growth optimization. We expect that MTSQDs have the potential to be of sub-nm scale spectral nonuniformity.

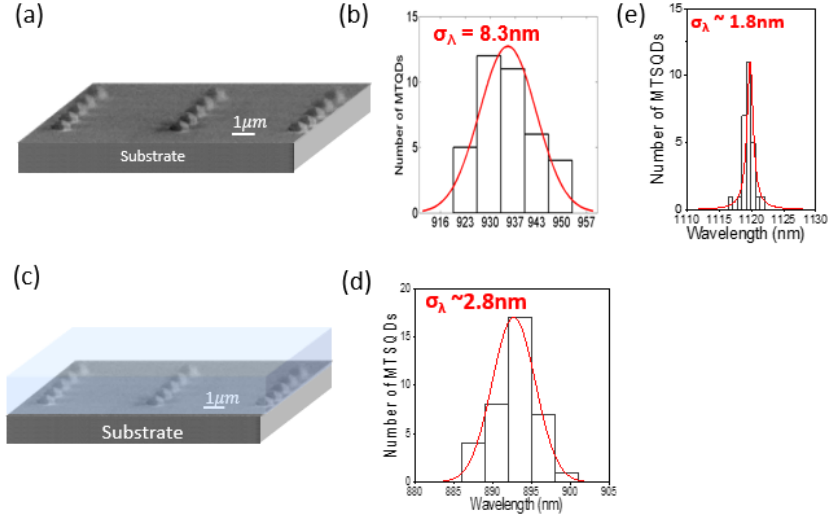

**Figure S2.** (a) SEM image of as-grown non-planarized 5×8 array of 4.25ML $In_{0.5}Ga_{0.5}As$ MTSQDs. (b) Histogram of emission wavelength of the 40 non-planarized $In_{0.5}Ga_{0.5}As$ MTSQDs in the array showing a spectral non-uniformity of 8.3nm (data from Ref. S12). (c) Schematic of the planarized MTSQDs array. (d) Histogram of emission wavelength of the 40 planarized $In_{0.5}Ga_{0.5}As$ MTSQDs in the array showing a spectral non-uniformity of 2.8nm (same as Fig. 1(e)). (e) Histogram of emission wavelength of the 40 non-planarized InAs MTSQDs in the array showing a spectral non-uniformity of 1.8nm (data from Ref. S13).

## 2. Optical Setup for Resonant measurement

All single photon emission characteristics data reported in the manuscript has been measured using resonant excitation scheme. With the sample containing the MTSQDs mounted in the cryostat, the measurements are done in vertical excitation and vertical detection geometry as shown in Fig. S3. The MTSQDs are excited resonantly at their neutral exciton emission wavelengths using Ti:Sa Mode lock laser with pulses of 3ps width. The unwanted scattered laser light into the detector is filtered out using a cross-polarization configuration (using two polarizers and one polarizing beam splitter) as shown in Fig. S3. The exciting laser electric field is along the [110] direction with an accuracy of ±3° and the photons with polarization along [-1 1 0] direction are detected, also with accuracy of ±3°. A cross-polarized extinction ratio $>1\times10^7$ is established for the resonant excitation studies reported here. The collected photons from the MTSQDs are spectrally resolved with a spectrometer with 15μeV resolution and detected by superconducting nanowire detectors.

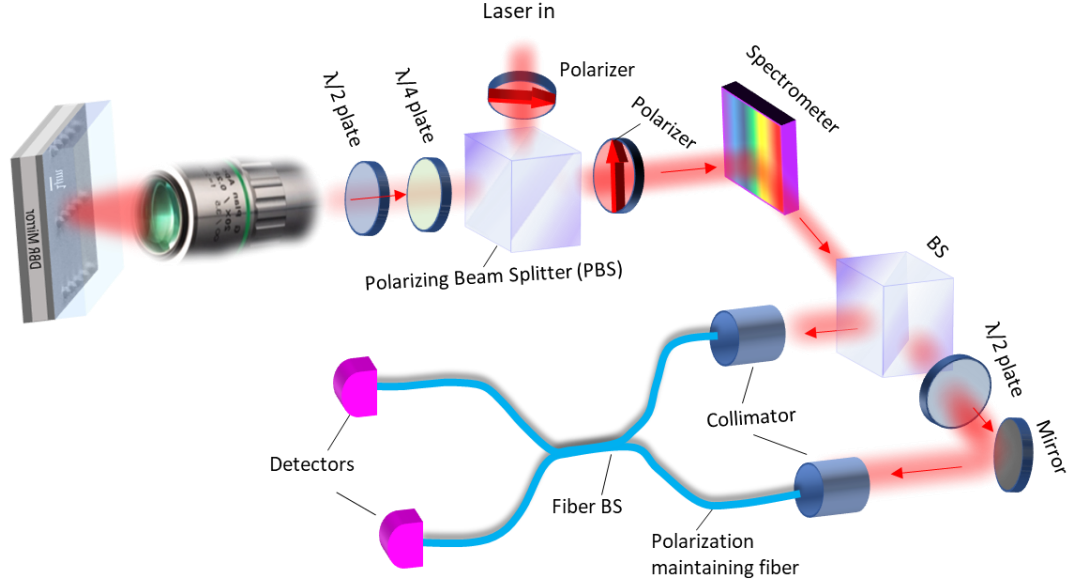

**Figure S3.** Schematic of the measurement instrumentation including the Hong-Ou-Mandel interferometry for resonant excitation where the scattered excitation laser light is filtered out with a cross-polarization configuration.

For the study of indistinguishability, we use the Ti-Sa laser to generate pairs of excitation pulses of width ~3ps with a time separation Δt=2ns, controlled by an unbalanced Michelson interferometer built on the laser side. The impinging laser is also polarized along [110]. The emitted photons from the MTSQD are passed through the polarizer to select emission polarized along [1-10] and are fed through the spectrometer to the first 50/50 beam splitter of the Hong-Ou-Mandel interferometer (Fig. S3). Photons passing through the beamsplitter are coupled to polarization maintaining fibers through two collimators at the transmitted and reflected ports of the beamsplitter. The path length difference is designed to match the time delay (2ns) of the excitation pulse. Photons passed through the fibers interfere at the fiber-inline-beam splitter (50/50) and detected by the supercomputing nanowire detectors. A λ/2-waveplate allows for rotating the polarization of photons in one arm of the interferometer with respect to the other arm in order to make the photons distinguishable on purpose for reference measurement. The instrument response function for the HOM measurement is ~50-100ps.

## 3. Photon collection efficiency and MTSQD quantum efficiency

To enhance the collection efficiency of photons emitted from the MTSQDs, we have grown a planarized 5×8 array of 4.25ML $In_{0.5}Ga_{0.5}As$ MTSQD on DBR (schematic shown in Fig. 1(c)).

The cross-section of the as-grown sample is shown in Fig. S4 (a) with the QD a marked as red dot. The typical PL from one of the MTSQDs sitting on the DBR is shown in Fig. S4(b) measured under above-gap excitation condition (50% of saturation power, with 640nm excitation) at 19.5K. To evaluate the enhancement of photon collection efficiency coming from bottom DBR mirror, we have grown a 4.25ML In$_{0.5}$Ga$_{0.5}$As MTSQD sample sitting on GaAs without DBR as comparison. We have measured PL from MTSQD on a DBR (Fig. S4(b)) and MTSQD on GaAs substrate (Fig. S4(c)) as comparative measurement to establish the enhancement of collection efficiency with DBR mirror. PL from a MTSQD without DBR (panel a, black curve) and a MTSQD with DBR (panel b) underneath are measured with the same optical setup and under same normalized power (50% of saturation power, with 640nm excitation) at 19.5K. The photon counts at detector is improved by a factor of ~10. The designed DBR has a reflectivity >95% in the range of 880-930nm. The sample structure (Fig. S4(a)) is designed to enhance the photon collection efficiency at the first objective lens by ~10, bringing it to ~12%. Fig. S4 (d) shows finite element method simulation of collection efficiency at the first objective lens. The measured PL signal enhancement is consistent with the simulated results, suggesting a factor of 10 enhancement in collection efficiency (the ratio of $\frac{\#\ of\ photon\ collected\ by\ first\ objective}{\#\ of\ photons\ emitted\ from\ MTSQD}$).

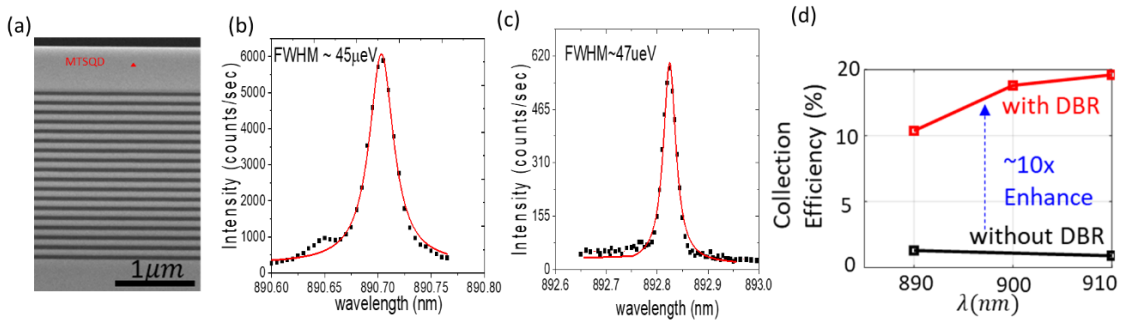

**Figure S4.** PL from a typical MTSQD sitting on DBR (panel (b), SEM of sample in panel (a)) and a typical MTSQD sitting on the GaAs (panel (c)) measured with 640nm excitation at 50% of saturation power at 19.5K. The finite element method simulation of collection efficiency at the first objective lens- indicating 10× enhancement by the DBR is shown in panel (d).

To quantitatively access the quantum efficiency of the as-grown MTSQD (Fig. 1(c) and Fig. S4(a)), we have measured the PL from the MTSQDs under resonant excitation and studied its emission as a function of power under resonant excitation. Fig. S5 shows the PL from one typical MTSQDs and the power dependence of the emission as a function of excitation power. At π pulse, one exciton is created inside MTSQDs per pulse. The detected photon count is 17K/sec. For estimation of the internal quantum efficiency, we calibrated the detection efficiency of the setup (shown in Fig. S3). We have used the laser tuned to the QD emission wavelength (~890nm) to calibrate the detection efficiency of the setup (the ratio of $\frac{Detection\ counts\ at\ detector}{\#\ of\ photons\ collected\ by\ the\ first\ objective}$). The detection efficiency of the setup (Fig. S3) is found to be ~$(1.81\pm0.02)\times10^{-3}$. Given the photon counts at detector being ~17K/sec at π pulse (Fig. 2(a)), the total number of photon collected by the first objective is ~$0.94\times10^7$/sec. Accounting for the 78MHz repetition rate of the laser and the 12% collection efficiency of emitted photons, we estimate internal quantum efficiency ~100% of the MTSQD.

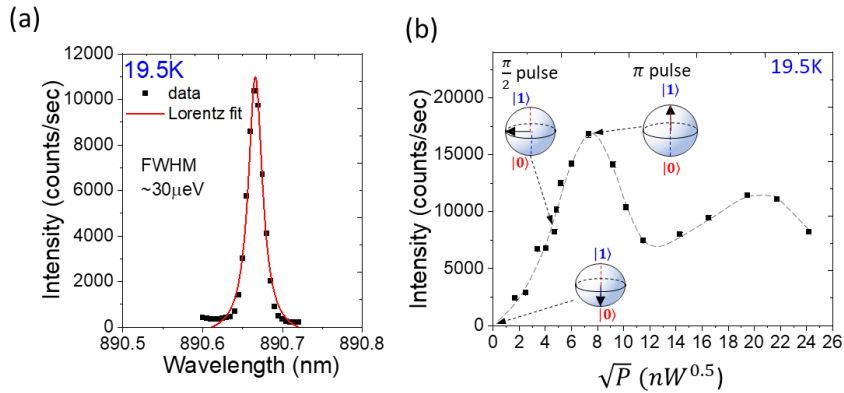

**Figure S5.** (a) Photoluminescence emission spectra from one typical MTSQD excited under resonant excitation with π/2 pulse, 19.6nW (1.6W/cm$^2$) at 19.5K. The red curve is a Lorentzian fit showing a measured linewidth of 30μeV. (b) Power dependent behavior of peak intensity vs the square root of laser power (proportional to excitation pulse area) showing clear Rabi oscillation. Error of the measured intensity is within the symbol size. The inserted Bloch sphere represents the switching from ground state |0⟩ (empty QD) to one exciton state |1⟩ (Fig. 2(a), recaptured here for easy reference).

### 4. Single Photon Purity

To assess the general behavior of single photon emission purity of the MTSQDs under resonant excitation, we have shown in Fig. 1(g) measured second-order correlation function of the photon emission from one MTSQD (2,2) (marked by the row and column number) in the array at 19.5K. Here we show results from one other MTSQD (MTSQD (4,2) in the array) in Figure S6 below. By calculating the ratio of the area under the τ=0 peak and the average of areas under the τ≠0 peaks, we obtain a $g^{(2)}(0)$ of 0.05 for the MTSQD, giving single photon emission purity ~97.5%. The measured single photon purity is consistent with previously reported behavior of MTSQDs[S12,S13]. A calculated curve based on theory[S14] is shown as the red line in Fig. 1(g) and Fig. S6. The number of photons emitted from QD ($n$) follows $n(t) \sim \exp\left(-\frac{t}{\tau_{QD}}\right)$ where $\tau_{QD}$ is the PL decay lifetime of the QDs. Thus the second order correlation function $g^{(2)}(\tau)$ is of the form (around each peak), $g^{(2)}(\tau) = <n(t)n(t+\tau)> \sim \exp(-|\tau - t_{exc}|/\tau_{QD})$. The coincidence count histogram is of the form $h(t) = A \int IRF(t)[\sum_{m \neq 0} \exp(-|\tau - t - mT|/\tau_{QD}) + g^{(2)}(0) \exp(-|\tau - t|/\tau_{QD})]dt$, where $IRF(t)$ is the instrument response function, $T$ is the time interval between excitation pulses, and m is an integer sequence number for an individual excitation pulse. The $g^{(2)}(0)$ extracted from theory is consistent with that extracted based on peak area ratios.

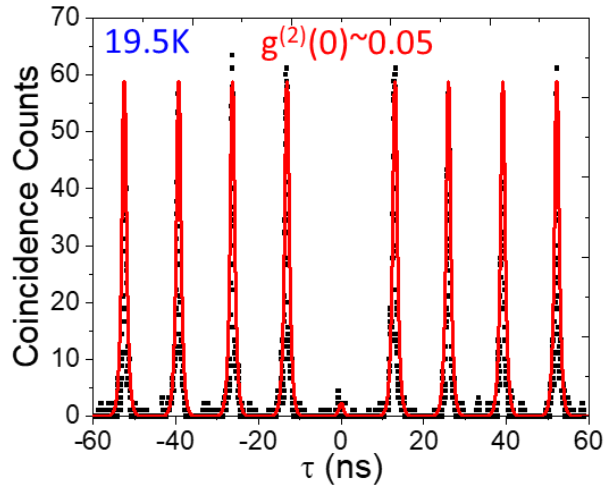

**Figure S6.** Histogram of coincidence counts of emission from MTSQD (4,2) using HBT setup with resonant excitation and with π/2 pulse at 19.5K. The red curve is the calculated curve based on theory[S14].

5. **Decay Dynamics:**

MTSQD (2, 2) (shown in Fig. 1 and 2 of manuscript) provides single photons in coherent superposition of two finely split state with Δ~6.4µeV (as shown in the beating signal in time-resolved PL of Fig. 2(b)). The observed beating pattern in time-resolved PL is also observed in other MTSQDs. Fig. S7 shows the measured TRPL data on MTSQD (4,2) at 19.5K under resonant excitation with π/2 pulse. It also shows beating pattern and reveals $\Delta = 3.8\mu eV$ and $T_1^{(a)} = T_1^{(b)} = 0.55ns$ from fitting (red line, Fig. S7) using Eq. 1. These data suggests that (1) MTSQDs have fine structure splitting <10ueV and (2) the emitted single photons are in coherent superposition of the states from the two finely split states.

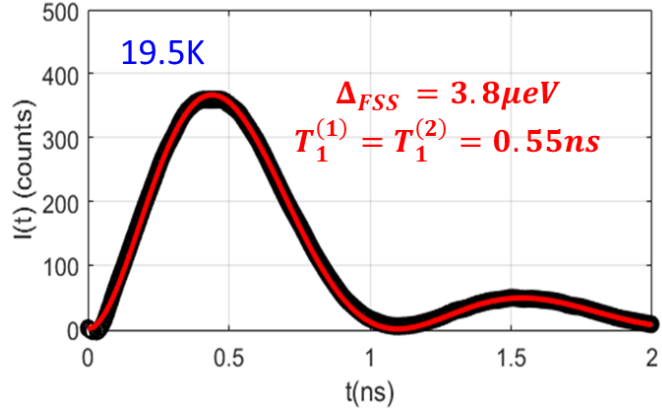

**Figure S7.** Measured time-resolved fluorescence from the MTSQD (4,2) excited under resonant excitation with π/2 pulse at 19.5K. The red curve is fitting to the data using Eq. 1 from our three-level model.

## 6. Indistinguishability: Effect of Phonon and Spectral Diffusion.

The effect of phonon and spectral diffusion on TPI visibility is analyzed following the model reported in Ref. S15 using Markovian approximation in addressing the phonon induced dephasing time. The TPI visibility is expressed[S15] as $V(T) = \frac{\Gamma}{\Gamma_{SD}+\gamma(T)+\Gamma}$ where $\Gamma = 1/2T_1$,

$\gamma(T) = \frac{\gamma_0}{\exp\left(\frac{\alpha}{T}\right)-1}\left[\frac{1}{\exp\left(\frac{\alpha}{T}\right)-1}+1\right]$ is the phonon induced dephasing and $\Gamma_{SD}$ is the spectral diffusion-induced dephasing time. We used the reported $\gamma_0$ and $\alpha$ from Ref. S15 to fit to our measured data shown in Fig. 3(a). Given the measured TPI visibility data both at 19.5K and 11.5K, the fitted behavior of visibility as a function of temperature is shown in Fig. S8 (black line in Fig. S8, measured data shown as red dots). The result matches the typical known temperature dependence of the exciton dephasing time in InGaAs/GaAs material system reported in Ref. S15 and indicates an expected visibility ~90% at 4K.

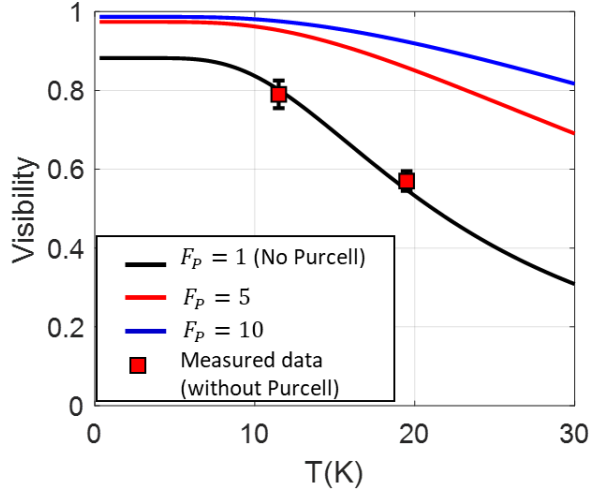

**Figure S8.** Measured visibility (red squares) and fit (black curve) for temperature dependent visibility without any Purcell enhancement. The red curve and the blue curve indicate expected visibility when a Purcell enhancement ~5 and ~10, respectively, are introduced.

The visibility can be further enhanced by embedding the MTSQD in appropriate photonic cavity structures introducing a Purcell enhancement[S16] - thus shortening the radiative lifetime $T_1$. With a Purcell enhancement $F_P$, the expected TPI visibility can be expressed as $V(T) = \frac{\Gamma \times F_P}{\Gamma_{SD}+\gamma(T)+\Gamma \times F_P}$. In Figure S8, we show the expected Visibility for a Purcell enhancement ~5 (red curve) and ~10 (blue curve) -typically achieved via DBR micropillar[S16], photonic crystal membrane[S17], or dielectric nanoantenna[S18] structures. At 4K, TPI visibility larger than 97% is estimated for Purcell enhancement 5, indicating that near unity indistinguishability can be readily reached for the MTSQD SPSs.

## 7. Three-level model

In this section we provide additional detailed information on the analysis of the photoluminescence, time resolved emission, single photon self-interference, and two-photon interference (TPI) of photons emitted from the MTSQDs based on the model of three-level structure of the exciton manifold. The content is organized as follows:

- In Section 7.a, we define the MTSQD as a three-level system and define a simple Hamiltonian that captures the resonant excitation and detection system. With this model, we derive (a) the power dependent PL intensity (Rabi oscillation) and (b) the expression of wavepacket of the emitted photon including the beating effect from the two fine structure split (FSS) states to compare with the data observed in time resolved photoluminescence measurement.

- In Section 7.b, exploiting the photon wavepacket derived in section 7.a and adapting the approach of Bylander et. al[S19] we show the analysis of outcome of photon self-interference and two photon Hong-Ou-Mandel interference to interpret the experimental measurements presented in this paper.

**7.a. Three-level Model of Dynamical Evolution of Exciton in MTSQD SPS:**

The measured time resolved photoluminescence of the MTSQDs as shown in Fig. 2(b) of the main text indicates a three-level structure of the ground level exciton of the MTSQD. In this section we analyze resonant fluorescence behavior of such a system shown schematically in Fig. S9(a). The two FSS split exciton states are denoted as $|1_a\rangle$ and $|1_b\rangle$ with energy separation $\Delta$, and the ground state (no exciton) is denoted as $|0\rangle$. Past studies have shown[S20] that the transition dipole moments of these two fine structure split states are linearly polarized at an angle ~$20^0$-$30^0$ with respect to the crystallographic direction [110] and [-1 1 0]. Here in this analysis we depict this angle as $\phi_0$ – as indiated in Fig. S9(b).

With the above assumption, we express the Hamiltonian corresponding to resonant excitation and detection measurement with the excitation E-field polarization along the [1 1 0] (zinc-blende) direction, and emitted photons polarized in the [-110] direction being collected into detection optics [Fig. S9(c)]. The overall Hamiltonian is expressed as,

$$H = H_{MTSQD} + H_{Laser-MTSQD} + H_{Cav} + H_{QD-Cav} + H_{Cav-Det} \quad (S1)$$

Here, the first term, Hamiltonian of the exciton manifold of the MTSQD by itself is,

$$H_{MTSQD} = \left(\omega_a - \frac{i}{2T_1^{(a)}} + F(t)\right)\sigma_a^\dagger \sigma_a + \left(\omega_b - \frac{i}{2T_1^{(b)}} + F(t)\right)\sigma_b^\dagger \sigma_b \quad (S2)$$

in which $\omega_a$ and $\omega_b$ represent the energy of the exciton states $|1_a\rangle$ and $|1_b\rangle$, respectively, and the effect of the radiative decay is captured as the introduced non-Hermiticity through decay times, $T_1^{(a)}$ and $T_1^{(b)}$. For simplicity, $\hbar$ is normalized to unity. $F(t)$ is the random energy shift fluctuating with time $t$ introduced by phonons and spectral diffusion, and is modelled by Langevin approach with Markovian approximation assuming a memoryless thermal reservoir[S19]. Thus we have $\langle F(t_1)F(t_2)\rangle = T_2^*\delta(t_1 - t_2)$, where $T_2^*$ is the dephasing time. Here $\langle\cdot\rangle$ denotes statistical avarage.

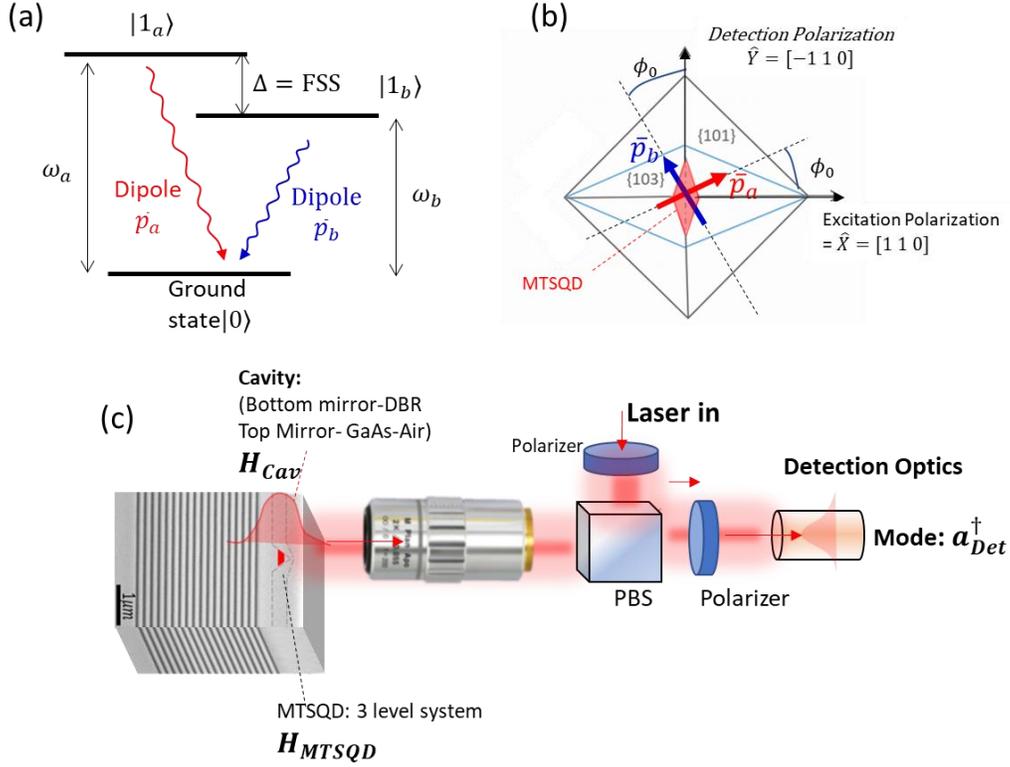

**Figure S9**. (a) Schematic of the three level structure of the MTSQD. (b) The orientations of the transition dipoles of the two exciton states with respect to the [110] type direction taken as an angle $\phi_0$. (c) The measurement geometry indicating the different components of the Hamiltonian as shown in equation (S1).

We use the $\bar{p}\cdot\bar{E}$ type interaction to express in equation (S1) the interaction term $H_{Laser-MTSQD}$ between the excitation laser and the exciton states. Under rotating wave approximation,

$$H_{Laser-QD} = p_0\,E_0\,e^{-\frac{2(t-t_0)^2\ln 2}{\Delta t_{Laser}^2}}\left[\cos\phi_0\,\sigma_a^\dagger e^{-i\omega_{Exc}t} + \sin\phi_0\,\sigma_b^\dagger e^{-i\omega_{Exc}t} + h.c.\right] \qquad (S3)$$

Here $E_0$ is the peak E-field strength (polarized along the [1 1 0] direction) of the excitation laser. $\Delta t_{Laser}$ is the laser pulse width, ~3ps for our case, and $\omega_{Exc}$ is the center wavelength of the pulsed laser. The transition dipole moments of the two exciton states are assumed to have equal amplitude $p_0$.

The emitted photons from the MTSQD are coupled to the cavity formed by the planar DBR bottom mirror and the GaAs-Air interface at the top. Photons from this DBR cavity are quickly leaked into the detection optics mode. This photon collection process is captured using the last three terms of the Hamiltonian in equation (S1) as,

$$H_{Cav} = (\omega_{Cav} - i\kappa_{Cav}) a_{Cav}^\dagger a_{Cav} \tag{S4}$$

$$H_{QD-Cav} = g\left[\sin\phi_0 \sigma_a^\dagger e^{-i\omega_{Cav}t} - \cos\phi_0 \sigma_2^\dagger e^{-i\omega_{Cav}t} + h.c.\right] \tag{S5}$$

$$H_{Cav-Det} = \kappa_{Cav}\, a_{Det}^\dagger a_{Cav} \tag{S6}$$

Here $\kappa_{Cav}$ is the rate of photon leakage from the DBR-GaAs cavity to the detection optics. Transforming to the rotating frame given by $H_0 = \omega_a \sigma_a^\dagger \sigma_a + \omega_b \sigma_b^\dagger \sigma_b + \omega_{Cav} a_{Cav}^\dagger a_{Cav}$, the Hamiltonian (sum of (S2) to (S6)) is simplified to $H = H_0 + e^{-iH_0 t}\, H_I\, e^{iH_0 t}$, where,

$$H_I(t) = p_0 E_0\, e^{-\frac{2(t-t_0)^2 \ln 2}{\Delta t_{Laser}^2}} \left[\cos\phi_0\, \sigma_a^\dagger e^{i\omega_a - i\omega_{Exc}t} + \sin\phi_0\, \sigma_b^\dagger e^{i\omega_b - i\omega_{Exc}t} + h.c\right]$$

$$+ g\left[\sin\phi_0 \sigma_a^\dagger e^{i\omega_a - i\omega_{Cav}t} - \cos\phi_0 \sigma_2^\dagger e^{i\omega_b - i\omega_{Cav}t} + h.c.\right]$$

$$+\kappa_{Cav} a_{Det}^\dagger a_{Cav}\, e^{-i\omega_{Cav}t} + F(t)\sigma_a^\dagger \sigma_a + F(t)\sigma_b^\dagger \sigma_b - i\kappa_{Cav} a_{Cav}^\dagger a_{Cav} - \frac{i}{2T_1^{(a)}}\sigma_a^\dagger \sigma_a - \frac{i}{2T_1^{(b)}}\sigma_b^\dagger \sigma_b \tag{S7}$$

The evolution of the combined state of the MTSQD exciton and the emitted photon under resonant excitation therefore follows,

$$i\frac{d}{dt}|\Psi_I(t)\rangle = H_I(t)|\Psi_I(t)\rangle \tag{S8}$$

with the state of the system, in the rotating frame, expressed as,

$$|\Psi_I(t)\rangle = c_0(t)|0\rangle + c_a(t)\sigma_a^\dagger|0\rangle + c_b(t)\sigma_b^\dagger|0\rangle + c_{cav}(t)a_{Cav}^\dagger|0\rangle + c_{Det}(t)a_{Det}^\dagger|0\rangle \tag{S9}$$

Equation (S8) and (S9) provide us the basic framework to analyze the resonant excitation measurements of PL from the MTSQD modelled as the three-level system. Specifically, this framework is now used to analyze the power-dependent and time-dependent PL response as follows:

**Power Dependent Resonant PL- Rabi Oscillation**

In the PL measurements, the QD exciton is first initialized by a short (~3ps) excitation laser pulse. Let us denote time t = 0 as the time at which the exciton state has just been initialized. Using the Hamiltonian in eq. (S7), and under the assumption that the excitation pulse is much shorter (~3ps) compared to the decay timescale (~350ps), we get the state of the system at the end of the excitation pulse to be,

$$|\Psi_I(0)\rangle = \sin\left(\int_{Pulse} \Omega(t)dt\right)[\cos\phi_0 \sigma_a^\dagger + \sin\phi_0 \sigma_b^\dagger]|0\rangle \tag{S10}$$

where $\Omega(t) = p_0 E_0 e^{-\frac{2(t-t_0)^2 \ln 2}{\Delta t_{Laser}^2}}$. Note that both the states $|1_a\rangle$ and $|1_b\rangle$ are populated by the laser since the excitation E-field has components along the transition dipole moment of both these exciton states (Fig. S9). For a Gaussian laser pulse of repetition rate $F_{Laser}$, full width half maximum is $\Delta t_{Laser}$, laser spot area = $A_{Spot}$, and the average incident power on the GaAs surface $P_{inc}$ we estimate,

$$\int_{Pulse} \Omega(t)dt = p_0 \sqrt{\frac{P_{inc}}{F_{Laser}} \frac{T_{GaAs}\eta_{GaAs}}{A_{Spot}} \frac{2\sqrt{\pi}}{\ln 2} \Delta t_{Laser}} = C\, p_0 \sqrt{P_{inc}} \tag{S11}$$

Here $F_{Laser}$ is the laser pulse rate ~78MHz for us; $\eta_{GaAs}$ is electromagnetic impedance of GaAs ~ 110 Ohm; $T_{GaAs}$ is the transmittance of the air-GaAs interface ~ 66%. $A_{Spot}$ is the laser spot area~$1.22\mu m^2$. The effect of these constants is combined into $C = \sqrt{\frac{T_{GaAs}\eta_{GaAs}}{F_{Laser}A_{Spot}} \frac{2\sqrt{\pi}}{\ln 2} \Delta t_{Laser}}$ which is a constant that only depends on the measurement instrumentation. $P_{Inc}$ is the average excitation power impinging on the sample surface. Equation S10 and S11 indicate the Rabi oscillation of the amplitude of the exciton states with respect to $P_{Inc}$ via the proportionality,

$$Intensity \propto \sin^2\left(C\, p_0 \sqrt{P_{inc}}\right) \tag{S12}$$

This oscillatory behavior manifests itself in the power dependent PL intensity data shown in Fig. 2(a). Note, in this analysis we did not take into account the excitation induced dephasing process[S21] – which results in the decay in the oscillatory amplitude of the power dependent PL as seen in Fig. 2(a) of the main text. Nevertheless, comparing the measured oscillatory period of the power-dependent PL emission with equation (S12), we estimate that the transition dipole moment of the exciton states to be of magnitude $p_0 \sim 70$ Debye.

## Time-Dependent Resonant PL: Description of Emitted Photon Wave Packet

Once the excitons in states $|1_a\rangle$ and $|1_b\rangle$ are populated by the excitation laser pulse to the state $|\Psi_I(0)\rangle$ given in equation (S10), the process of single photon emission is analyzed using the same Hamiltonian as in equation (S7). Under the expansion shown in equation (S9), applying the Hamiltonian in equation (S7) and Schrodinger equation in eq. (S8), we have,

$$i\frac{d}{dt}c_0(t) = 0 \tag{S13}$$

$$i\frac{d}{dt}c_a(t) = \left[-\frac{i}{2T_1^{(a)}} + F(t)\right]c_a(t) + g\sin\phi_0\, e^{-i\omega_{Cav}t+i\omega_a t}\, c_{cav}(t) \tag{S14}$$

$$i\frac{d}{dt}c_b(t) = \left[-\frac{i}{2T_1^{(b)}} + F(t)\right]c_b(t) - g\cos\phi_0\, e^{-i\omega_{Cav}t+i\omega_b t}\, c_{cav}(t) \tag{S15}$$

$$i\frac{d}{dt}c_{cav}(t) = -i\kappa_{Cav}c_{cav}(t) + g\sin\phi_o\, e^{i\omega_{Cav}t-i\omega_a t}\, c_a(t) - g\cos\phi_0\, e^{i\omega_{Cav}t-i\omega_b t}\, c_b(t) \tag{S16}$$

and,

$$i\frac{d}{dt}c_{Det}(t) = \kappa_{Cav}e^{-i\omega_{Cav}t}c_{Cav}(t) \tag{S17}$$

The evolution can be further simplified by applying the weak coupling limit. The QD-cavity coupling strength g in our planar DBR cavity is $\sim 1\mu eV$ - much smaller than the decay rate $\kappa_{Cav} \sim 10$meV. Thus, $\kappa_{Cav} \gg g$ which leads to $|c_{Cav}(t)| \ll |c_a(t)|, |c_b(t)|$. Under this weak coupling limit, solving equations (S13) to (S16), we have

$$c_a(t) \approx e^{-\frac{t}{2T_1^{(a)}}} e^{-i\int_0^t F(t)dt}\, c_a(0) \tag{S18}$$

and,

$$c_b(t) \approx e^{-\frac{t}{2T_1^{(b)}}} e^{-i\int_0^t F(t)dt} c_b(0) \tag{S19}$$

The integrated effect of the random energy fluctuation $F(t)$ is reflected as a randomized additional phase $\phi(t) = \int_0^t F(t)dt$ that represents the dephasing process. From Langevin formulation[S19] based on a memoryless reservoir we have

$$\langle e^{-i\phi(t_1)} e^{i\phi(t_2)} \rangle = e^{-\frac{|t_1-t_2|}{T_2^*}} \tag{S20}$$

Finally, the amplitude of the photon state collected in the detection optics mode can be expressed, from equation (S16) to (S17), as,

$$c_{Det}(t) = g \int_0^t e^{-i\omega_{Cav}t}[\sin\phi_0 \, e^{-i\omega_a t} c_a(t) - \cos\phi_0 e^{-i\omega_b t} c_b(t)]dt \tag{S21}$$

Using equation (S18) and (S19), respectively, as the expressions for $c_a(t)$ and $c_b(t)$, and using equation (S10) for the state at t = 0, we get the emitted photon wavepacket collected in the detection optics expressed as,

$$|\Psi_{Photon}(t)\rangle = \int_0^t f(t) a_{Det}^\dagger(t)|0\rangle \, dt \tag{S22}$$

Where $f(t) = g\sin\phi_0 \cos\phi_0 \sin(\int_{Pulse} \Omega(t)dt) \left[ e^{-i\omega_a t - \frac{t}{2T_1^{(a)}}} - e^{-i\omega_b t - \frac{t}{2T_1^{(b)}}} \right] e^{-i\phi(t)}$.

equation (S22), we can express the time resolved intensity as,

$$I(t) \propto |f(t)|^2 \propto \left| e^{-i\Delta t} e^{-\frac{t}{2T_1^{(a)}}} - e^{-\frac{t}{2T_1^{(b)}}} \right|^2 \tag{S23}$$

This analytical form is shown in equation (1) of the main text and is used to fit the measured time resolved PL data as shown in Fig. 2(b) of the main text and Fig. S7 to extract $\Delta$, $T_1^{(a)}$, and $T_1^{(b)}$.

### 7.b Photon Interference Measurements:

The photon wavepacket expression derived in last section is now exploited to derive the expected outcome of single photon self-interference and two-photon interference, as follows:

**Self-Interference of the emitted photon - Fringe Contrast:**

Self-interference is exploited in the coherence time measurement, where the photon wavepacket emitted into the detection optics is split into two branches and then recombined into one branch with a relative delay of $\tau_D$, as indicated in Fig. S10.

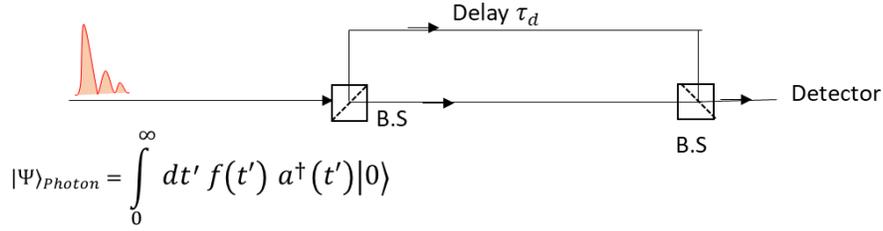

$$|\Psi\rangle_{Photon} = \int_0^\infty dt'\, f(t')\, a^\dagger(t')|0\rangle$$

**Figure S10**. Schematic of photon correlation measurement set-up to determine self-interference of a single photon.

Using the photon wavepacket expression in equation (S22), the resultant wavepacket at the detector can be expressed as,

$$|\Psi_{at\ detector}(t)\rangle \propto \int_0^t f(t) a^\dagger_{Det}(t)|0\rangle dt + \int_0^t f(t+\tau_D) a^\dagger_{Det}(t)|0\rangle dt \qquad (S24)$$

Thus, intensity at the detector can be expressed in the following proportionality relation:

$$I_{Det} \propto \int_0^\infty |f(t)|^2\, dt + Re\left[\int_0^\infty f(t)f^*(t+\tau_d) dt\right] \qquad (S25)$$

The fringe contrast is expressed as,

$$Fringe\ contrast = \frac{\max(I_{Det}) - \min(I_{Det})}{\max(I_{Det}) + \min(I_{Det})} = \frac{\left|\int_0^\infty f(t)f^*(t+\tau_d) dt\right|}{\int_0^\infty |f(t)|^2 dt} \qquad (S26)$$

$$= \frac{1}{I_0}\left[\int_0^\infty dt\, e^{-\frac{t}{T_1}} \sin\left(\frac{\Delta}{2} t\right) \sin\left(\frac{\Delta}{2}(t+\tau_d)\right)\right] e^{-\frac{\tau_d}{2T_1}} \langle e^{-i\phi(t)+i\phi(t+\tau_d)} \rangle \qquad (S27)$$

$$= \frac{1}{I_0}\left[\int_0^\infty dt\, e^{-\frac{t}{T_1}} \sin\left(\frac{\Delta}{2}t\right) \sin\left(\frac{\Delta}{2}(t+\tau_d)\right)\right] e^{-\frac{\tau_d}{2T_1}-\frac{\tau_d}{T_2^*}} \quad (S28)$$

Here $I_0 = \int_0^\infty |f(t)|^2\, dt = \frac{\Delta^2 T_1^3}{2(1+\Delta^2 T_1^2)}$ is the intensity of a single photon wavepacket. Equation (S28) provides the analytical expression for the fringe contrast that is shown in equation (2) of the main text. This equation is used to fit the data in Fig. 2(c).

**Two Photon Interference- Hong-Ou Mandel Configuration:**

In the two photon HOM interference measurement, two different photon wavepackets emitted from the same MTSQD with a time difference of $\Delta t \approx 2ns$ are collected into the detection optical branch. One of these two photons is delayed, also by $\Delta t$ to compensate for the emission time difference and then interfered with the other photon. Schematically this is shown in Fig. S11.

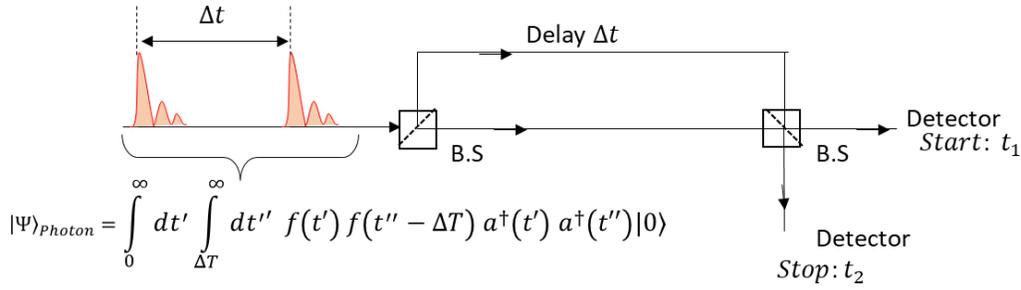

**Figure S11**. A schematic representation of the HOM two-photon interference measurement.

We adapt the approach of Bylander[S19] and apply it for the single photon wave packet shown in equation (S22) to derive the analytical expression of the two-photon correlation function as,

$g^{(2)}(t_1, t_2)$

$= |f(t_1 - \Delta T)f(t_2 - 2\Delta T)|^2 + |f(t_2 - \Delta T)f(t_1 - 2\Delta T)|^2 + |f(t_1)f(t_2 - 2\Delta T)|^2$

$+ |f(t_2)f(t_1 - 2\Delta T)|^2 + |f(t_1)f(t_2 - \Delta T)|^2 + |f(t_2)f(t_1 - \Delta T)|^2$

$+ |f(t_1 - \Delta T)f(t_2 - \Delta T)|^2 \left[2 - \left(\langle e^{-i\phi_1(t_1)+i\phi_1(t_2)-i\phi_2(t_2)+i\phi_2(t_1)}\rangle + c.c.\right)\right]$ (S29)

The photon interference effect is contained in the time average of the random phase fluctuation, $\phi_1(t)$ and $\phi_2(t)$, of the first and the second photon emitted from the same QD, separated by ~2ns.

Since 2ns is much longer than the dephasing timescale at the operating temperature of 19.5K, it is reasonable to assume that $\phi_1(t)$ and $\phi_2(t)$ are uncorrelated [S19]. Thus,

$$\langle e^{-i\phi_1(t_1)+i\phi_1(t_2)-i\phi_2(t_2)+i\phi_2(t_1)} \rangle = \langle e^{-i\phi_1(t_1)+i\phi_1(t_2)} \rangle \langle e^{-i\phi_2(t_2)+i\phi_2(t_1)} \rangle = e^{-\frac{2|t_1-t_2|}{T_2^*}} = e^{-\frac{2|\tau|}{T_2^*}} \quad (S30)$$

The time resolved HOM $g^{(2)}(t_1, t_2)$ as shown in equation (S29) is plotted in Fig. 2(f) of the main text along with the measured result. Further, from equation (S29) and (S30), integrating over $t_2$- the detection time of the "stop" detector, the analytical expression of the HOM $g^{(2)}(\tau)$ coincidence count histogram can be derived as,

$$g_{\parallel}^{(2)}(\tau) = \int_0^\infty dt\, e^{-\frac{2t}{T_1}} \sin^2\left(\frac{\Delta}{2}t\right) \sin^2\left(\frac{\Delta}{2}(t+|\tau|)\right) \left[1 - e^{-\frac{2|\tau|}{T_2^*}}\right] e^{-\frac{|\tau|}{T_1}} \quad (S31)$$

For perpendicular polarization, i.e., no interference at the second HOM beam splitter, the center peak of the HOM coincidence count takes the form,

$$g_{\perp}^{(2)}(\tau) = \int_0^\infty dt\, e^{-\frac{2t}{T_1}} \sin^2\left(\frac{\Delta}{2}t\right) \sin^2\left(\frac{\Delta}{2}(t+|\tau|)\right) e^{-\frac{|\tau|}{T_1}} \quad (S32)$$

These functional forms in equation (S31) and (S32) are used to fit to the measured HOM $g^{(2)}$ data shown in Fig. 2 of the main text and to extract the dephasing time $T_2^*$ of the exciton state of the MTSQDs.

**References:**

(S1) Michler, P.; Quantum Dots for Quantum Information Technologies. Springer, **2017**.

(S2) Guha, S,; Madhukar, A.; Rajkumar, K. C. Onset of incoherency and defect introduction in the initial stages of molecular beam epitaxical growth of highly strained $In_xGa_{1-x}As$ on GaAs(100), *Appl. Phys. Lett.* **1990**, 57, 2110.

(S3) Xie, Q.; Chen, P.; Kalburge, A.; Ramachandran, T.R.; Nayfonov, A.; Konkar, A.; Madhukar, A. Realization of optically active strained InAs island quantum boxes on GaAs (100) via molecular beam epitaxy and the role of island induced strain fields. *J. Cryst. Growth.* **1995**, 150, 357-363.

(S4) Kim, J.; Aghaeimeibodi, A.; Richardson, C. J. K.; Leavitt, R. P.; Waks, E. Super-radiant emission from quantum dots in a nanophotonic waveguide. *Nano Lett.* **2018**, 18, 8, 4734-4740.

(S5) Gidney, C.; E. Martin, How to factor 2048 bit RSA integers in 8 hours using 20 million noisy qubits. *Quantum* **2021**, 5, 433.

(S6) Kiravittaya, S.; Rastelli, A.; Schmidt, O. G. Advanced quantum dot configurations. *Rep. Prog. Phys.* **2009**, 72, 046502.

(S7) Rigal, B.; Jarlov, C.; Gallo, P.; Dwir, B.; Rudra, A.; Calic, M.; Kapon, E. Site-controlled quantum dots coupled to a photonic crystal molecule. *App. Phys. Lett.* **2015**, 107, 141103.

(S8) Chen, Y.; Zadeh, I. E.; Jons, K. D.; Fognini, A.; Reimer, M. E.; Zhang, J.; Dalacu, D.; Poole, P. J.; Ding, F.; Zwiller, V.; Schmidt, O. G. Controlling the exciton energy of a nanowire quantum dot by strain fields. *Appl. Phys. Lett.* **2016**, 108, 182103.

(S9) Madhukar, A. Growth of semiconductor heterostructures on patterned substrates: defect reduction and nanostructures. *Thin Solid Films*. **1993**, 231, 8.

(S10) Konkar, A.; Rajkumar, K.C.; Xie, Q.; Chen, P.; Madhukar, A.; Lin, H.T; Rich, D.H. In-situ fabrication of three-dimensionally confined GaAs and InAs volumes via growth on non-planar patterned GaAs(001) substrates. *J. Cryst. Growth*. **1995,** 150, 311.

(S11) Konkar, A.; "Unstrained and strained semiconductor nanostructure fabrication via molecular beam epitaxical growth on non-planar patterned gallium arsenide(001) substrate" PhD Dissertation, University of Southern California, 1999.

(S12) Zhang, J.; Chattaraj, S.; Lu, S.; Madhukar, A. Mesa-top quantum dot single photon emitter arrays: growth, optical characteristics, and the simulated optical response of integrated dielectric nanoantenna-waveguide systems. *Jour. App. Phys.* **2016**, 120, 243103.

(S13) Zhang, J.; Huang, Q.; Jordao, L.; Chattaraj, S.; Lu S.; Madhukar, A. Planarized spatially-regular arrays of spectrally uniform single quantum dots as on-chip single photon sources for quantum optical circuits. APL photonics 2020, 5, 116106.